\documentclass{mn2e}

\input psfig

\title[Outward migration of extrasolar planets]
	{Migration of extrasolar planets to large orbital radii}
	
\author[D. Veras \& P.J. Armitage]
       {Dimitri Veras$^{1,2}$\thanks{email: {\tt Dimitri.Veras@colorado.edu}} and 
       Philip J. Armitage$^{1,2}$ \\
	$^1$JILA, University of Colorado, 440 UCB, Boulder CO 80309-0440, USA \\
	$^2$Department of Astrophysical and Planetary Sciences, University of Colorado, 
	    Boulder CO 80309-0391, USA}	

\begin{document}

\maketitle

\begin{abstract}
Observations of structure in circumstellar debris discs provide circumstantial
evidence for the presence of massive planets at large (several tens of AU) orbital
radii, where the timescale for planet formation via core accretion
is prohibitively long. Here, we investigate whether a population
of distant planets can be produced via outward migration
subsequent to formation in the inner disc. Two possibilities
for significant outward migration are identified. First, cores that form early
at radii $a \sim 10 \ {\rm AU}$ can be carried to larger radii via gravitational
interaction with the gaseous disc. This process is efficient if there is strong
mass loss from the disc -- either within a cluster or due to photoevaporation
from a star more massive than the Sun -- but does not require the extremely 
destructive environment found, for example, in the core of the Orion Nebula. 
We find that, depending upon the disc model, gas disc migration can 
yield massive planets (several Jupiter masses) at radii of
around 20-50~AU. Second, interactions within multiple planet systems
can drive the outer planet into a large, normally highly eccentric orbit.
A series of scattering experiments suggests that this process is most
efficient for lower mass planets within systems of unequal mass ratio. This
mechanism is a good candidate for explaining the origin of relatively low mass
giant planets in eccentric orbits at large radii.
\end{abstract}


\section{Introduction}

With one exception (Konacki et al. 2003), all confirmed
extrasolar planets have been discovered by the Doppler velocity technique.  The
selection effects inherent to radial velocity surveys (Cumming, Marcy \& Butler 1999)
favor the detection of planets at small orbital radii.  To date, about half of
the known planets have semi-major axis $a < 1$ AU, while the most distant - 55 Cnc d - lies
at $5.9$ AU from its parent star.
\footnote{From the online Extrasolar Planets Encyclopedia, at http://cfa-www.harvard.edu/planets/catalog.html
(Schneider 2003) as of March 26th, 2003.}.
Indirect evidence, however, suggests that there could be a sizable population of massive planets
at much greater radii.  Recent observations of dusty debris
around Vega have been interpreted as suggesting the presence of a planet of a few
Jupiter masses with $a > 30$ (Wilner et al. 2002).  Further, simulations modeling
circumstellar dust discs suggest a planet lies at a distance of $55-65$ AU from
Epsilon Eridani (Ozernoy et al. 2000).

Forming planets in these outer locations is difficult.  Gas giants must
form before the disc dissipates, at timescales no greater than $5-10$ Myr
(Haisch, Lada, \& Lada 2001).  In standard core accretion models (Safranov 1969),
the timescale for building the core of a giant planet increases rapidly with
radius, with a $t_{form}$ scaling approximately as $a^2$ (Pollack et al. 1996).
Although such models are undoubtedly oversimplified (Pollack et al. 1996;
Bryden, Lin \& Ida 2000), it is hard to avoid the conclusion that forming
massive planets at radii of several tens of AU within $10$ Myr is difficult.
Indeed, this has led to the suggestion that Uranus and Neptune may have
formed at smaller radii in our own Solar System (Thommes, Duncan \& Levison 1999, 2002).   
Motivated by these issues, we investigate the possibility of forming
massive planets at small $a$, followed by outward migration. In Sections 2 and 3, 
we consider sequentially the two mechanisms that have been extensively studied in the 
context of inward migration: planet-disc
interactions (Goldreich \& Tremaine 1980; Lin \& Papaloizou 1986; Lin, Bodenheimer
\& Richardson 1996; Trilling et al. 1998) and gravitational scattering after disc
dissipation (Rasio \& Ford 1996; Weidenschilling \& Marzari 1996; Lin \& Ida 1997;
Ford, Havlickova \& Rasio 2001; Terquem \& Papaloizou 2002). Our conclusions are 
briefly summarized in Section 4.

\section{Migration via gas disc interactions}

\subsection{Methods}

We calculate the orbital evolution of massive planets embedded 
within an evolving protoplanetary disc using a variant of the 
approach described by Armitage et al. (2002). We use a simple, 
one-dimensional (i.e. vertically averaged) treatment to model 
the evolution of a protoplanetary disc evolving under the 
action of both internal viscous torques and external torques 
from one or more embedded planets (Goldreich \& Tremaine 1980; 
Lin \& Papaloizou 1986; Trilling et al. 1998; Trilling, Lunine \& 
Benz 2002). For a disc with surface density $\Sigma(R,t)$, the 
governing equation is,
\begin{equation}
  { {\partial \Sigma} \over {\partial t} } = 
 { 1 \over R } { \partial \over {\partial R} } 
 \left[ 3 R^{1/2} { \partial \over {\partial R} } 
 \left( \nu \Sigma R^{1/2} \right) - \nonumber \\
 { { 2 \Lambda \Sigma R^{3/2} } \over 
 { (G M_*)^{1/2} } } \right] + \dot{\Sigma}_{w}.
\label{eqsigma} 
\end{equation}   
Here, $\nu$ is the kinematic viscosity which models 
angular momentum transport within the disc gas, and 
$\dot{\Sigma}_w$ is a term which allows for mass to be 
lost from the disc -- for example as a consequence of 
photoevaporation. The second term within the 
brackets describes how 
the disc responds to the planetary torque, $\Lambda (R, a)$, 
where this function is the rate of angular momentum 
transfer per unit mass from the planet to the disc. 
For a planet in a circular orbit at radius $a$, 
we take, 
\begin{eqnarray} 
 \Lambda & = & - { {q^2 G M_*} \over {2 R} } 
 \left( {R \over \Delta_p} \right)^4 \, \, \, \, R < a \nonumber \\
 \Lambda & = & { {q^2 G M_*} \over {2 R} } 
 \left( {a \over \Delta_p} \right)^4 \, \, \, \, R > a 
\label{eqtorque} 
\end{eqnarray} 
where $q = M_p / M_*$, the mass ratio between 
the planet and the star, 
\begin{equation}
 \Delta_p = {\rm max} ( H, \vert R - a \vert ),
\end{equation} 
and $H$ is the scale height of the disc. Guided by 
detailed protoplanetary disc models (Bell et al. 1997), 
we adopt $H = 0.05 R$.

The transfer of angular momentum leads to orbital 
migration of the planet at a rate,
\begin{equation}
 { { {\rm d} a } \over { {\rm d} t } } = 
 - \left( { a \over {GM_*} } \right)^{1/2} 
 \left( { {4 \pi} \over M_p } \right) 
 \int_{R_{\rm in}}^{R_{\rm out}} 
 R \Lambda \Sigma {\rm d} R,
\end{equation} 
if the only torque on the planet comes from the 
gravitational interaction with the disc.

In the core accretion model for giant planet formation 
(e.g. Pollack et al. 1996), the accretion of the gaseous 
envelope is predicted to take longer than any other 
phase of the formation process (several Myr in the 
baseline model for Jupiter presented by Pollack et al. 1996). 
In particular, the time scale for accretion is much 
longer than the time scale on which a sufficiently massive 
planet can establish a gap in the protoplanetary disc, since  
numerical simulations show that an approximately 
steady-state gap can be set up by a massive planet 
within $\sim 10^2$ orbital periods (e.g. Lubow, Seibert \& 
Artymowicz 1999). A consequence of the inequality of these 
time scales is that massive planets -- those of several Jupiter masses -- 
probably accrete most of their envelopes 
{\em subsequent} to the development of a gap in the 
protoplanetary disc. Numerical simulations show how 
this accretion may occur. Gas from the outer disc 
penetrates the leaky tidal barrier created by the 
planet, and flows inward to form a small circumplanetary 
disc around the growing planet (Lubow, Seibert \& Artymowicz 1999; D'Angelo, Henning
\& Kley 2002; Bate et al. 2003).

The existence of mass flow across gaps onto planets is 
intrinsically a two (or three) dimensional phenomenon 
(e.g. the discussion in Artymowicz \& Lubow 1996). The 
torque function (Eq. \ref{eqtorque}) used in our 
one dimensional code establishes a clean gap for all 
planet masses above about $0.1 M_J$, and this gap 
precludes any mass flow across the gap, or onto the 
planet. To allow for the mass growth of planets, we 
have therefore modified the one dimensional treatment 
to explicitly include mass flow from the outer disc 
on to the planet. We begin by making an approximate fit to the 
results of two-dimensional numerical simulations (Lubow et al. 1999; 
D'Angelo et al. 2002). We define the {\em efficiency} 
of mass accretion across the gap via a parameter $\epsilon$, 
which is the planetary accretion rate as a fraction of the 
disc accretion rate at large radii (away from the location 
of the planet). The results of the aforementioned numerical 
simulations can then be approximated by the formula,
\begin{equation} 
 { \epsilon \over \epsilon_{\rm max} } \simeq 
 1.668 \left( { M_p \over M_J} \right)^{1/3} 
 e^{- {M_p \over {1.5 M_J} } } + 0.04,
\end{equation} 
where $M_J$ is the mass of Jupiter and $\epsilon_{\rm max}$ 
is an adjustable parameter which can be used to test how 
the results depend upon the overall efficiency of 
planetary accretion. We use the 
above equation to calculate at each timestep the 
appropriate planetary accretion rate. We then 
remove the required amount of mass from the first 
zone on the outer edge of the gap, and add it to 
the mass of the planet. Note that we assume that all 
the mass flow onto the planet originates from the outer disc, 
and do not permit any material to `bypass' the planet 
and flow directly from the outer disc to the inner disc.

Mass accretion across the gap onto a planet may also 
be expected to lead to accretion of angular momentum, 
given that the specific angular momentum of gas at 
the outer gap edge exceeds that of the planet. This 
effect -- which it is easy to show can have a significant 
influence on the migration rate -- cannot be straightforwardly 
measured from existing numerical simulations\footnote{Bate et al. 
(2003), for example, explicitly exclude this advective torque 
from their estimates of the migration rate, due to difficulties 
in measuring the net torque from the small scale circumplanetary 
disc formed in their simulations.}. For this paper, we 
adopt the simplest approach, and assume that the accreted 
gas has the same specific angular momentum as gas at a 
radius $R_{gap} = 1.6 a$. This fixed radius approximates 
the location of the outer edge of the gap throughout 
most of the calculation.

Eq. (\ref{eqsigma}) is solved on a fixed, non-uniform 
mesh using standard explicit numerical methods (e.g. Pringle, 
Verbunt \& Wade 1986). The mesh is uniform in a scaled 
variable $X \propto \sqrt{R}$. Typically, 300 grid points 
are used, with an inner boundary at 0.1 AU and an outer 
boundary at 200 AU. A zero-torque ($\Sigma = 0$) boundary 
condition is applied at $R_{in}$. For the protoplanetary 
disc model adopted, the outer boundary is at sufficiently 
large radius that the choice of boundary condition there 
has no influence on the results.

\subsection{Protoplanetary disc model}

We model the protoplanetary disc as a viscous accretion 
disc (Lynden-Bell \& Pringle 1974) suffering mass loss 
at large radius as a consequence of photoevaporation 
(e.g. Shu, Johnstone \& Hollenbach 1993). Motivation 
for considering models of this form is provided 
first and foremost by observations of photoevaporative 
flows in Orion (Johnstone, Hollenbach \& Bally 1998), 
and is discussed further by Clarke, Gendrin \& 
Sotomayor (2001), Matsuyama, Johnstone \& Hartmann (2003), 
and Armitage, Clarke \& Palla (2003). Briefly, we write the viscosity as a 
fixed (in time) power-law in radius. For most of the calculations we have 
adopted the form, 
\begin{equation}
 \nu = 3 \times 10^{13} \left( {R \over {1 \ {\rm AU}}} \right) \ {\rm cm^2 \ s^{-1}}.
\end{equation} 
This yields a steady-state surface density profile $\Sigma \propto R^{-1}$, 
which is similar to that derived from more detailed protoplanetary 
disc models over the radii of interest (Bell et al. 1997). To test 
how sensitive the conclusions are to this assumption, we have also 
run one model with a different form for the viscosity,
\begin{equation}
 \nu = 1.3 \times 10^{13} \left( {R \over {1 \ {\rm AU}}} \right)^{3/2} \ {\rm cm^2 \ s^{-1}}.
\end{equation}
Mass loss from the disc scales with radius as,
\begin{eqnarray}
 \dot{\Sigma}_{w} & = & 0, \,\,\,\,\,\,\,\,\,\,\, R < R_g \nonumber \\
 \dot{\Sigma}_{w} & \propto & R^{-1}, \,\,\,\,\, R > R_g,
\label{eq_massloss}
\end{eqnarray}
with $R_g = 5 \ {\rm AU}$. We express the normalization of the mass 
loss via a parameter $\dot{M}_{\rm wind}$, which is defined as 
the total mass loss from the disc for a disc with an outer 
edge at 25 AU. The instantaneous rate of mass loss will therefore 
differ from this value depending upon the extent of the disc.

The initial surface density profile for the $\nu \propto R$ disc model is,
\begin{equation}
 \Sigma = \Sigma_0 \left( 1 - \sqrt{R_{in} \over R} \right) {1 \over R} 
 e^{-R / R_0}, 
\end{equation}
while the $\nu \propto R^{3/2}$ model is identical except for the 
replacement of $1/R$ by $1/R^{3/2}$. 
Here, $\Sigma_0$ is a constant used to define the initial accretion 
rate, while $R_0$ is a truncation radius which sets a smooth 
exponential cut off to the surface density at large radius. For 
$R \ll R_0$, a disc described by this initial condition has a 
constant accretion rate, so we specify the initial surface 
density of our models via this inner accretion rate $\dot{M}_{\rm init}$.

\subsection{Results}
\begin{figure}
\psfig{figure=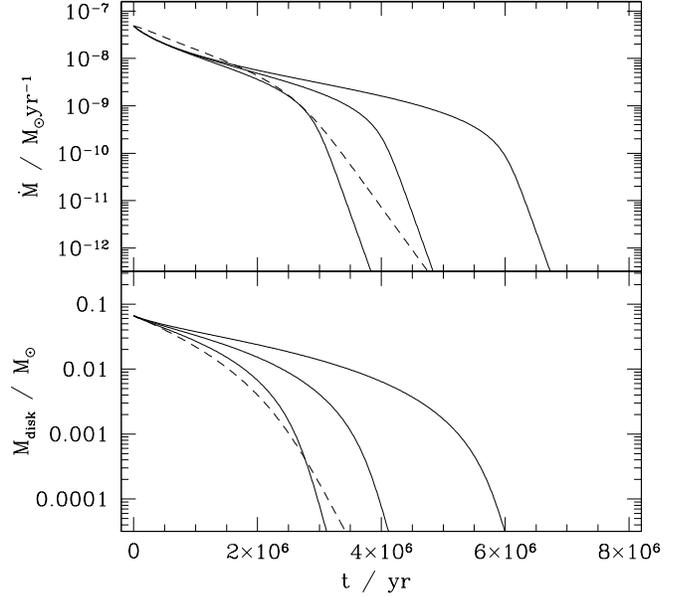,width=3.5truein,height=3.5truein}
\caption{Evolution of the accretion rate (upper panel) and mass (lower 
	panel) of the protoplanetary disc models used for migration 
	calculations. The solid curves show the evolution for  
	models with $\dot{M}_{\rm wind} = 10^{-9} \ M_\odot {\rm yr}^{-1}$,
	$\dot{M}_{\rm wind} = 2.5 \times 10^{-9} \ M_\odot {\rm yr}^{-1}$, and 
	$\dot{M}_{\rm wind} = 5 \times 10^{-9} \ M_\odot {\rm yr}^{-1}$ (with increasing 
	mass loss rates leading to smaller lifetimes). The dashed curve shows 
	a variant model with $\nu \propto R^{3/2}$ and $\dot{M}_{\rm wind} = 5 \times 
	10^{-9} \ M_\odot {\rm yr}^{-1}$. The other parameters of the models 
	are as described in the text.}
\label{fig_gas1}
\end{figure}

Fig. \ref{fig_gas1} shows how the evolution of the accretion rate and 
disc mass varies with the strength of photoevaporative mass loss. We 
have computed models with our standard viscosity ($\nu \propto R$) that 
have $\dot{M}_{\rm wind} = 10^{-9} \ M_\odot {\rm yr}^{-1}$, 
$\dot{M}_{\rm wind} = 2.5 \times 10^{-9} \ M_\odot {\rm yr}^{-1}$, and 
$\dot{M}_{\rm wind} = 5 \times 10^{-9} \ M_\odot {\rm yr}^{-1}$. 
For consistency with observational determinations of protoplanetary 
disc parameters in nearby star-forming regions (e.g. Gullbring et al. 1998), 
we adopt for all of these models an initial accretion rate of 
$\dot{M}_{\rm init} = 5 \times 10^{-8} \ 
M_\odot {\rm yr}^{-1}$, and a truncation radius of $R_0 = 10 \ {\rm AU}$. 
This yields an initial disc mass of $0.066 \ M_\odot$. An additional model 
(shown as the dashed curve in Fig. \ref{fig_gas1}) was calculated with 
the $\nu \propto R^{3/2}$ viscosity law and a mass loss of 
$\dot{M}_{\rm wind} = 5 \times 10^{-9} \ M_\odot {\rm yr}^{-1}$. 
With identical choices of $\dot{M}_{\rm init}$ and $R_0$, the 
initial disc mass for this model was $0.067 \ M_\odot$.

As expected from previous calculations (Clarke et al. 2001; Matsuyama et al. 2003), 
all four models show qualitatively similar evolution. There is 
an initial phase in which the disc mass and accretion rate decline 
slowly, due primarily to mass accretion onto the star. Subsequently, 
the mass and accretion rate drop more rapidly as the evolution 
becomes dominated by mass loss via the wind (Clarke, Gendrin \& 
Sotomayor 2001). Higher rates of mass loss reduce the 
disc lifetime, but all models have observationally acceptable 
lifetimes in the range between 4~Myr and 7~Myr. Most importantly for 
our purposes, in all four models mass loss is at least reasonably 
important (relative to accretion) for the overall evolution of the 
disc. The fraction of the initial disc mass that is lost in the wind 
varies between 40~percent and 52~percent for the $\nu \propto R$ models, 
and is 39~percent for the $\nu \propto R^{3/2}$ model. We note that 
the fraction lost in the wind is only a weak function of $\dot{M}_{\rm wind}$, 
because the interval of time over which the wind acts is reduced for 
higher instantaneous mass loss rates.

\begin{figure*}
\centerline{\psfig{figure=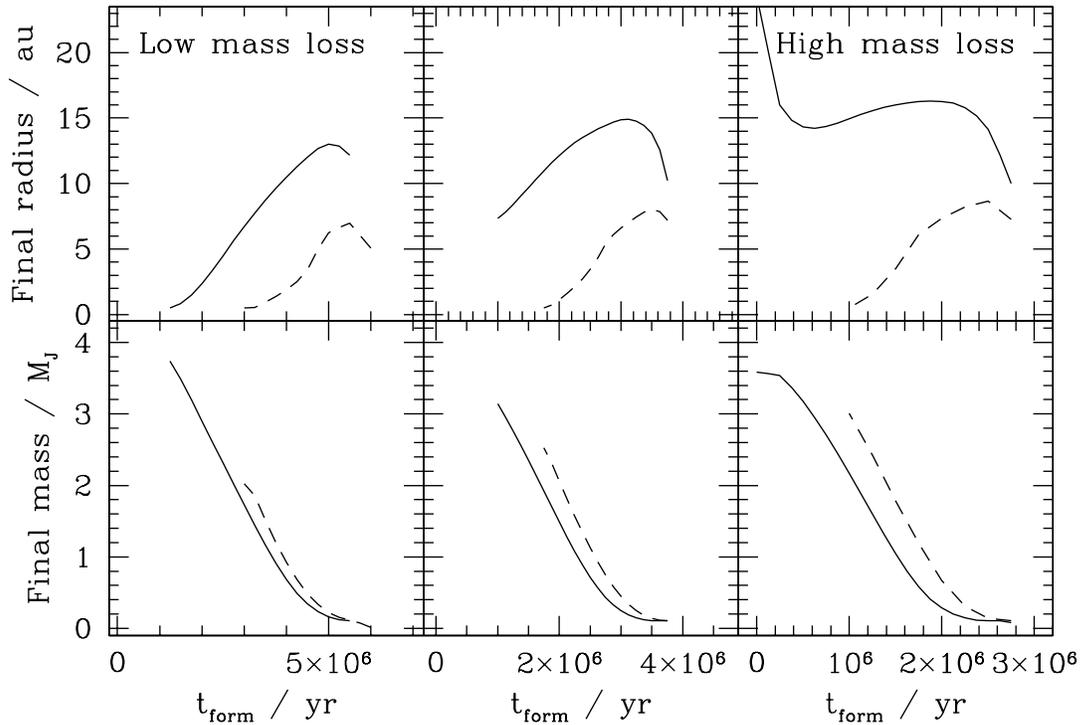,width=6truein,height=6truein}}
\vskip-1.8truein
\caption{The final orbital radius and final mass of planets following 
	disc migration, shown as a function of the planets' formation 
	epoch. The dashed and solid curves show results for planets formed 
	at initial orbital radii of 5~AU and 10~AU respectively. The 
	left-hand panels show the extent of migration in a model 
	disc with $\dot{M}_{\rm wind} = 10^{-9} \ M_\odot {\rm yr}^{-1}$, 
	the centre panels $\dot{M}_{\rm wind} = 2.5 \times 10^{-9} \ M_\odot {\rm yr}^{-1}$, 
	while the right-hand panels depict results from the 
	$\dot{M}_{\rm wind} = 5 \times 10^{-9} \ M_\odot {\rm yr}^{-1}$ model.}
\label{fig_gas2}
\end{figure*}

To investigate how planets migrate within the evolving disc, we run the 
protoplanetary disc models repeatedly. In each run, we 
add an initially low mass planet ($0.1 \ M_J$) into the disc at a
specified time and radius. We then allow the planet to grow and 
migrate within the evolving disc, and record the final planet 
mass and orbital radius once the disc has been either accreted 
or lost in the wind. By varying the formation time $t_{\rm form}$,  
and the formation radius $a_{\rm form}$, we study how the 
final outcome depends upon when and where in the disc massive 
planets form.

Fig. \ref{fig_gas2} shows the results for the disc with the 
standard viscosity law and varying rates of mass loss. 
We considered planet formation radii of 
5 AU and 10 AU, safely outside any estimate of the snow line 
(Sasselov \& Lecar 2000), and took $\epsilon_{\rm max} = 1$. 
The final planet mass and orbital 
radius are plotted as a function of the formation time. 
For the low and intermediate rates of mass loss, the sense 
of orbital migration is predominantly {\em inward}. Planets 
formed near the end of the disc lifetime end up in orbits 
close to where they formed, accrete relatively little disc 
gas, and remain as low mass objects. Planets formed earlier 
migrate inwards under the action of gravitational torques, 
and have time to grow to several Jupiter masses. These 
results are consistent with previous studies of migration 
(Trilling et al. 1998; Armitage et al. 2002; Trilling, 
Lunine \& Benz 2002). 

For the highest rate of mass loss from the disc, however,   
qualitatively different evolutionary tracks, shown 
in the right-hand panels of Fig. \ref{fig_gas2}, are 
obtained. The enhanced mass loss means that there is a larger range of 
disc radii across which the radial velocity of the 
gas is outward. This can drive significant outward 
migration. For our choices of parameters, we find 
that inward migration persists for all planets 
formed at 5 AU, while outward migration is the 
rule for planets formed at 10 AU. For formation 
times between about 1~Myr and 2~Myr, migration 
approximately stalls (similar to the behaviour 
reported by Matsuyama, Johnstone \& Murray 2003, 
though for slightly different reasons), while 
for earlier formation times 
the gas can drive these outer planets to 
radii of around 20 AU or greater. Significant accretion onto 
the planet occurs throughout this time, so the 
planets stranded at larger radii are all  
predicted to be massive objects.

\begin{figure}
\psfig{figure=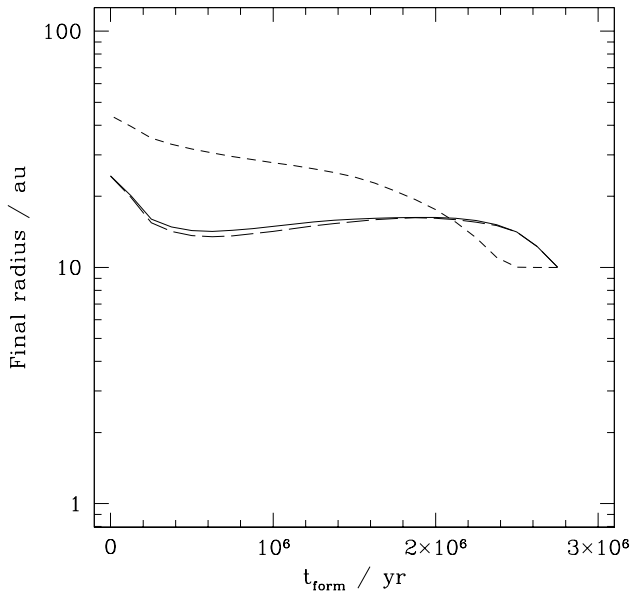,width=3.5truein,height=3.5truein}
\caption{Sensitivity of the migration results to changes in the model 
	parameters. The solid curve shows the extent of migration in 
	the standard disc model with $\nu \propto R$, $\epsilon_{\rm max} = 1$, 
	and $\dot{M}_{\rm wind} = 5 \times 10^{-9} \ M_\odot {\rm yr}^{-1}$. 
	The short dashed curve shows a model with a different viscosity law ($\nu \propto R^{3/2}$), 
	and the long dashed curve a model with less efficient accretion on to the 
	planet ($\epsilon_{\rm max} = 0.5$).}
\label{fig_gas3}
\end{figure}

Any attempt to distil the inherently multi-dimensional physics of planetary 
migration into a fast one dimensional scheme is bound to be approximate, and 
there are particularly obvious uncertainties in our models for planet growth 
and disc evolution. We have already demonstrated, as our main result, that for 
planets forming at radii of around 10~AU a switch between inward and outward 
migration occurs when $\dot{M}_{\rm wind}$ is varied by a factor of a few. 
The mass loss rate via photoevaporation is clearly a vital control parameter. 
To check how important some of the other parameters are, we have recalculated 
the migration of planets formed at 10~AU in two different models. In one, 
we altered the assumed disc viscosity (to $\nu \propto R^{3/2}$ rather 
than $\nu \propto R$), with the new viscosity chosen to produce a steady-state 
surface density profile $\Sigma \propto R^{-3/2}$. This scaling is one 
used often in studies of the Solar nebula (Weidenschilling 1977). A second 
model was computed with the standard viscosity law, but with an accretion 
efficiency parameter $\epsilon_{\rm max} = 0.5$. This change halves the 
rate of growth of planets via accretion. Both models used the high rate 
of mass loss previously found to be conducive to outward migration.

Fig.~\ref{fig_gas3} shows the final radii attained by planets in the 
three models. Outward migration occured in all three models, reflecting the 
primary importance of the assumed mass loss rate in determining the 
fates of the model planets. Changing the efficiency of accretion on to 
the planet made negligible difference to the final planetary radii 
(though it reduced the final masses of the model planets 
by approximately a factor of two). Substantially greater migration, however,  
was obtained in the calculation with the different viscosity law, 
despite the fact that a {\em smaller} fraction of the initial disc 
mass was actually lost to the wind in this case. We interpret this 
as being a side effect of the different surface density profile. 
Outward migration occurs when the torque from the inner disc 
exceeds that from the disc at radii beyond the planet's orbit. 
The steeper surface density profile of the $\Sigma \propto R^{-3/2}$ 
means that there is less mass initially exterior to the planet. 
As this mass is lost in the wind, the now unbalanced torque 
from the inner disc is more effective in driving the planet outward.

\def\mapright#1{\smash{
	\mathop{\rightarrow}\limits^{#1}}}
\def\mapdown#1{\downarrow\rlap
	{$\vcenter{\hbox{$\scriptstyle#1$}}$}}

\begin{table*}
\begin{center}
\def\temp#1{\multicolumn{1}{|r|}{#1}}
\tabcolsep=0.5mm
{\scriptsize\rm
\begin{tabular}{l|c|c|c|c|c|c|c|c|c|c|c|c|c|c|c|} \cline{2-16}
{$\enskip \enskip \enskip \enskip \mapright{} \mu_1$} \\
{$\matrix{ & \mapdown{}\cr & \mu_2 &}$}
& \rule{0mm}{2mm}1(-5)
& 2.5(-5)
& 5(-5)
& 7.5(-5)
& 1(-4)
& 2.5(-4)
& 5(-4)
& 7.5(-4)
& 1(-3)
& 2.5(-3)
& 5(-3)
& 7.5(-3)
& 1(-2)
& 2.5(-2)
& 5(-2)  \\ \hline
\temp{1(-4)} & \bf INNER & \bf INNER & \bf INNER & \bf BOTH & \bf BOTH & \bf OUTER & * & * & * & & & & & & \\ \hline
\temp{5(-4)} & & & * & * & \bf NONE & \bf BOTH & \bf BOTH & * & * & \bf NONE & \bf NONE & & & & \\ \hline
\temp{1(-3)} & & & & & \bf NONE & * & \bf OUTER & \bf OUTER & \bf NONE & \bf NONE & \bf NONE & * & \bf NONE & & \\ \hline
\temp{5(-3)} & & & & & & & \bf NONE & \bf NONE & \bf NONE & * & \bf NONE & \bf NONE & \bf NONE & \bf OUTER & \bf OUTER \\ \hline
\end{tabular}
}
\end{center}

\caption{Classifying the extent of migration or lack thereof for
different combinations of mass ratios, with initial parameters $0 < e_1 <  0.01$, 
$0 < e_2 < 0.01$, $0^{\circ} < i_1 < 5^{\circ}$, $0^{\circ} < i_2 < 5^{\circ}$,
randomly chosen initial orbital angles, and initial outer semimajor axis that lies within
the region of quasi-periodic orbits.  ``Inner'' implies that the inner planet exhibited
most of the outward migration, ``outer'' implies that the outer planet exhibited most of the outward
migration, ``both'' implies that both planets exhibited outward migration, ``none'' implies neither
planet exhibited significant migration, and ``*'' implies that $< 10\%$ of the 300 systems
run were stable.}

\end{table*}

\subsection{Observational implications}

What are the implications of our gas disc migration 
calculations for the origin of planets at large 
orbital radii?  We believe that three general conclusions 
can be drawn. First, 
mass loss from the outer disc can drive substantial 
outward migration, even when the mass loss is modest 
enough that the disc can survive for several Myr. 
Higher rates of mass loss would lead to more 
dramatic migration, but the resulting short disc lifetimes 
might preclude planet formation. In the central regions 
of the Orion Nebula, for example, Johnstone et al. (1998) 
infer mass loss rates between $2 \times 10^{-8} \ M_\odot 
{\rm yr}^{-1}$ and $6 \times 10^{-7} \ M_\odot 
{\rm yr}^{-1}$, with correspondingly small estimated 
disc lifetimes. Our results suggest that photoevaporation 
can have a major impact on planetary migration even 
in substantially more benign environments. 
Second, outward migration 
driven by the gas disc is favoured in systems where 
photoevaporative mass loss is stronger. The predicted 
rate of mass loss due to photoevaporation for a Solar 
mass star is rather small (Shu, Johnstone \& Hollenbach 1993), 
so for relatively isolated stars we would expect outward 
migration only for stars significantly more massive than the Sun. 
An alternative possibility is that the mass loss is driven 
by external irradiation, as in the case of Orion (Johnstone, 
Hollenbach \& Bally 1998). Finally, outward migration 
via this mechanism is a relatively slow process, which 
occurs on the viscous time of the protoplanetary disc. 
There is ample time for initially low mass planets 
to accrete substantial gaseous envelopes, so we would 
expect planets at large radius to be massive objects.

\section{Migration in multiple planet systems}

\subsection{Introduction}

If the initial outcome of the planet formation process is a system of several massive planets,
subsequent gravitational interactions can lead to many possible outcomes.  Planets
may collide, be ejected from the system, or settle into quasi-periodic or periodic orbits.
Theoretical investigations of the orbital evolution of two-planet systems (Ford, Havlickova,
\& Rasio 2001) and few-planet systems (Chambers, Wetherill, \& Boss 1996) have been conducted
with detailed descriptions of special cases, such as equal mass planets on coplanar orbits.

Gladman (1993) established that systems with two close planets exhibit chaotic but
quasi-periodic behavior given the appropriate initial conditions.  Ford, Havlickova, and
Rasio (2001) explored the evolution of such systems when the mass of each of two planets revolving
around a solar-type star in nearly circular, coplanar orbits is equal to $10^{-3} M_{\odot}$.  We
will expand on this study by considering different mass ratios for the two planets revolving around a
central star.  In doing so, we will show that outward migration {\it is} possible, for {\it either} planet.

\subsection{Methods}

\subsubsection{Motivation}

The setup for our set of simulations is motivated primarily by the gravitational scattering
experiments performed by Ford, Havlickova, \& Rasio (2001).  That study considered the
interaction between two planets of mass $10^{-3} M_{\odot}$ (on the order of a Jupiter mass)
and in an initially close configuration.  The resulting branching ratios of system
outcomes was explored.  Systems became unstable by ejecting a planet or by a collision,
one between planets or one between a planet and a star.   Stable systems
remained or settled into quasi-periodic orbits over $2$ Myr.

\begin{figure}
\centerline{\psfig{figure=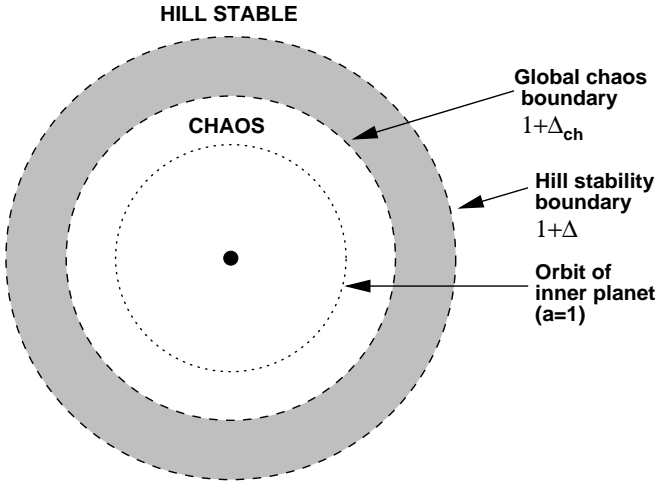,width=3.5truein,height=2.5truein}}

\caption{Schematic illustration of the initial conditions for scattering experiments. We randomly populate
orbits in the shaded region, which lie outside the global chaos boundary but are inside the boundary
defining guaranteed Hill stability.}
\label{dia}
\end{figure}

\begin{figure}
\centerline{\psfig{figure=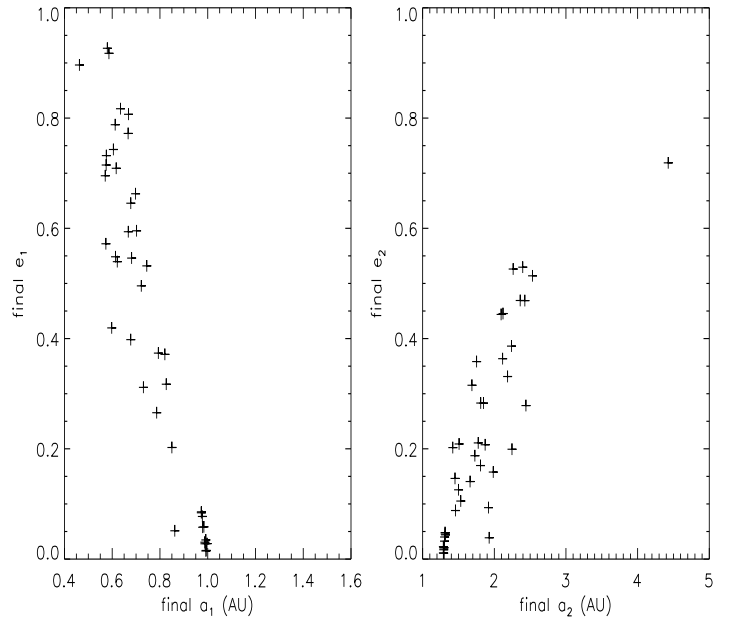,width=3.75truein,height=3.5truein}}

\caption{Scatter plots of the final eccentricity vs. semimajor axis for both planets in the
stable systems where $\mu_1 = 5 \times 10^{-4}, \ \mu_2 = 1 \times 10^{-3}$.}
\label{fig5}
\end{figure}

The initial configuration of both planets in Ford, Havlickova, \& Rasio's (2001) simulations
is motivated by the ``Hill Stability Criterion'', an analytic result first applied to planetary
systems by Gladman (1993).  A system is said to be Hill stable if the planets cannot approach each
other closely for all time.  We adopt Gladman's (1993) notation: $\Delta$ represents
the least separation for which both planets, {\it for sure}, will be Hill stable.
Nothing can be said about the Hill stability of a system for a
separation less than $\Delta$.  $\Delta_{ch}$ represents the greatest separation
at which ``global chaos" will occur.  A system that is ``globally chaotic'' might
produce collisions or ejections.  For a separation greater than
$\Delta_{ch}$, the planets {\it might} exhibit stable quasi-periodic orbits.
The geometry is illustrated in Fig. \ref{dia}.
Gladman's (1993) analytic second order expansion for small values of $\mu_1$ and $\mu_2$ yields
an approximate expression for $\Delta$,

$$\Delta \simeq
2\cdot 3^{\frac{1}{6}} {(\mu_{1} + \mu_{2})}^{\frac{1}{3}} +
{2\cdot 3^{\frac{1}{6}} {(\mu_{1}+\mu_{2})}^{\frac{2}{3}} - 
\frac{11\mu_{1} + 7\mu_{2}}{3^{\frac{11}{6}}{(\mu_{1} + \mu_{2})}^{\frac{1}{3}}}}.$$

Unlike $\Delta,$ an approximation for $\Delta_{ch}$ can only be found empirically.  By
developing overlap resonance criteria for two-planet systems, Wisdom
(1980) derived the following approximation when $\mu_{1} = \mu_{2} = \mu$:

$$ \Delta_{ch} \simeq 2{\mu}^{\frac{2}{7}}$$

Ford, Havlickova, \& Rasio (2001) ran simulations where the initial semimajor
axis separation of the planets lay between $\Delta$ and $\Delta_{ch}$, so that
the planets would neither be in a Hill Stable configuration nor become unstable
immediately.  In this work, we sample the entire initial separation range spanned
by $\Delta$ and $\Delta_{ch}$ in order to best detect planets that may migrate
outwards and remain on quasi-periodic orbits.

\subsubsection{Simulation Setup}

We denote the semimajor axis of the initially
inner planet $a_1$, the initially outer planet $a_2$, and the respective planet mass/central
mass ratios as $\mu_1$ and $\mu_2$.  Other orbital elements will distinguish the planets with
a subscript of ``1'' or ``2''.  All runs were performed with $a_1 = 1$ AU, so that the results
may be scaled easily to any ratio of semimajor axes, such as for $a_1 = 5 - 10$ AU.

All numerical runs of the three-body system were performed with a Bulirsch-Stoer routine
from the HNbody integration package (Rauch \& Hamilton 2002).  The routine ran for 2 Myr with an
accuracy parameter of $10^{-12}$, and was rerun with uniformly smaller initial timesteps when
necessary until this accuracy was achieved.  For each run, the Bulirsch-Stoer routine began
with an initial timestep of $0.05$ yr, and the
orbital elements of each body were output every $0.01$ Myr.  In most cases, energy and angular
momentum errors, expressed by $(E-E_{0})/{E_{0}}$ and $(L_{Z}-L_{Z_{0}})/L_{Z_{0}}$, where
$E_{0}$ is the initial total system energy and $L_{Z_{0}}$ is the initial total angular momentum
in the direction perpendicular to the orbital plane, did not exceed $10^{-7}$.  In the most 
pathological cases, energy and z-angular momentum errors were conserved
to within $10^{-4}$.  Angular momenta in the other two directions were typically conserved to
two order of magnitudes better than the z-angular momentum.

We are most interested in the evolution of stable systems in which both planets remain bound.  We
define stable systems as systems where both planets have semimajor axes which never exceed $10^{3}$ AU and 
eccentricities which never exceed $0.99$.  Further, we deemed a system unstable if at any time
the HNbody code outputted a negative value for the semimajor axis or eccentricity of either planet.

A preliminary exploration of phase space revealed that of the systems that become unstable, most
did so within 0.05 Myr.  Thus, each system underwent several checks at 0.05 Myr;
unstable systems were terminated, and stable systems were evolved for a total of 2 Myr.
In order to reduce computer time, we imposed additional conditions on systems after 0.05 Myr.
Systems not satisfying these conditions were terminated.  Given that $a_1 = 1$ AU for each
run, the conditions are

1) 1.2 AU $< a_2 <$ 3.0 AU,  

2) $a_1 < 1.05$ AU,  

3) If ${a_2}/{a_1} < 1.5$, then $e_2 < 0.1$,  

4) If ${a_2}/{a_1} > 1.5$, then $e_2 < 0.3$.  

\noindent{These} conditions were chosen based on a preliminary exploration of the properties of unstable systems.
For example, we found that after 0.05 Myr, if the semimajor axis of the outer planet is within 50\% of
$a_1$, then a high ($> 0.1$) eccentricity of the outer planet implies that the system will become unstable.

\begin{figure*}
\centerline{\psfig{figure=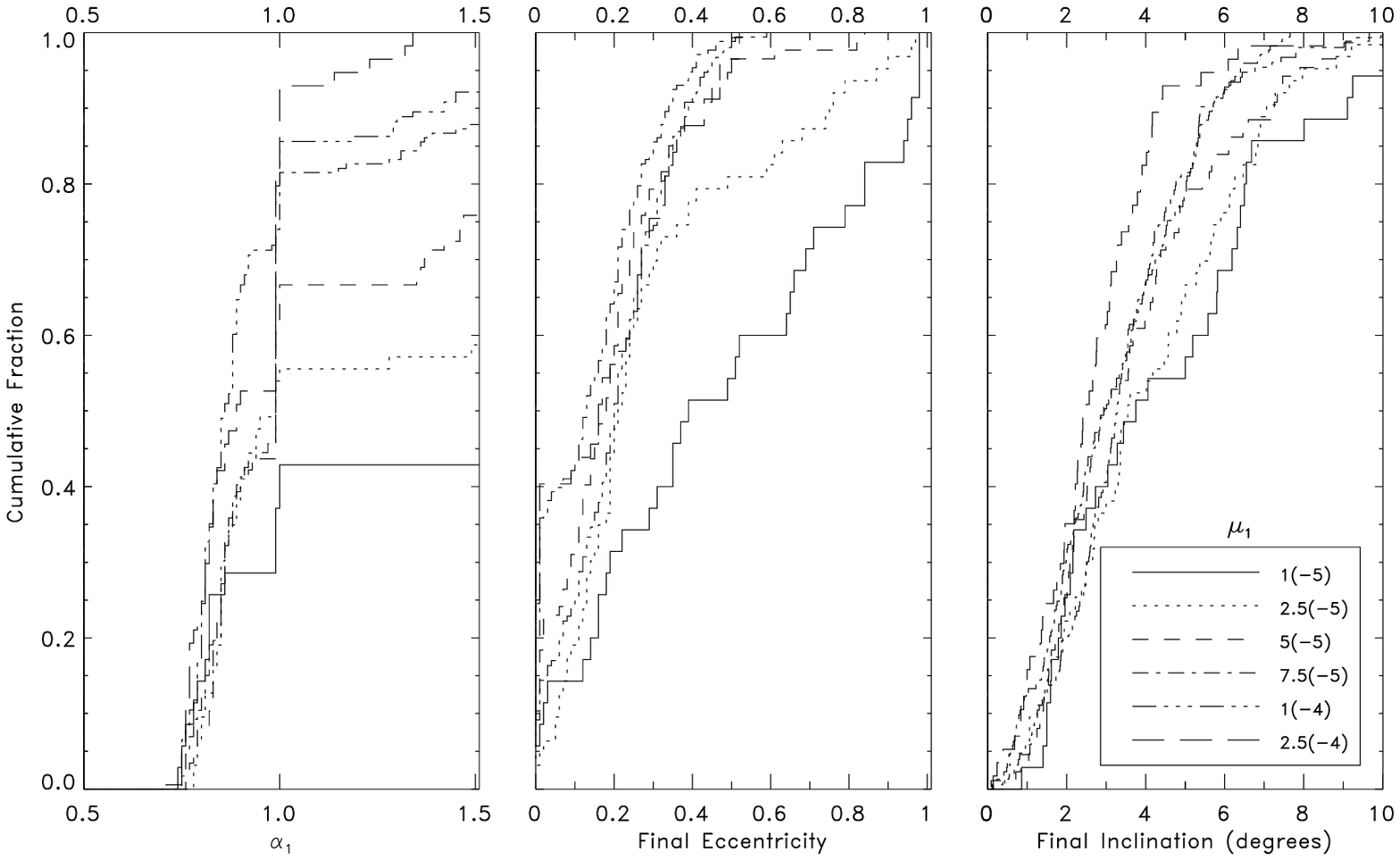,width=6.75truein,height=3.5truein}}
\vspace{-0.10truein}
\centerline{The Inner Planet}

\vbox{}
\vbox{}

\centerline{The Outer Planet}
\vspace{-0.35truein}
\centerline{\psfig{figure=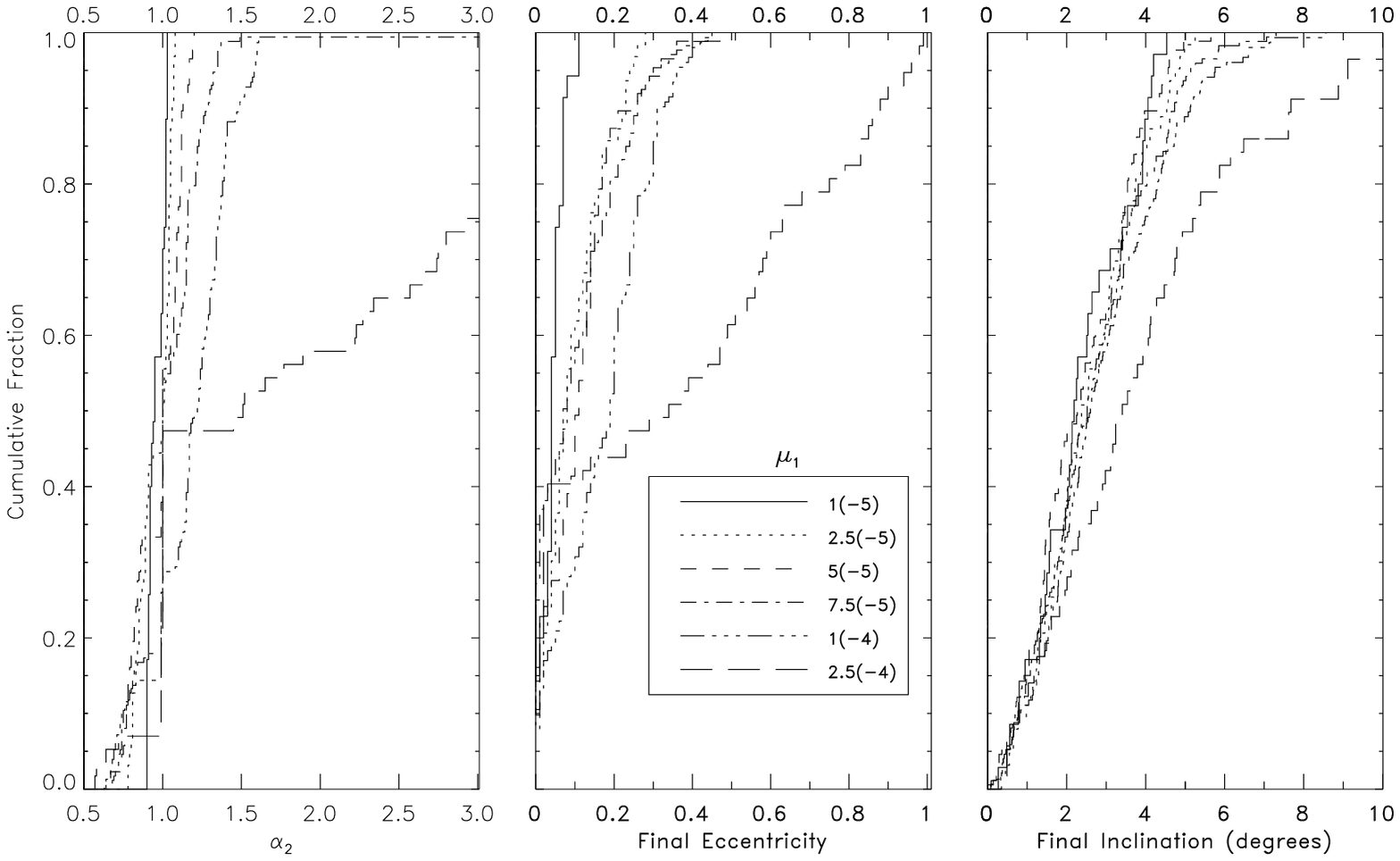,width=6.75truein,height=3.5truein}}

\caption{Cumulative probability distributions for stable systems with $\mu_2 = 1 \times 10^{-4}$.
For clarity, values of $\mu_1$ are abbreviated such that $1 \times 10^{-5} \equiv 1(-5)$.
The upper panel represents orbital parameter distributions of the inner planet, and the
lower panel the outer planet.}
\label{fig1}
\end{figure*}

By definition, long-term chaotic behavior may differ drastically due to an infinitesimal change
in initial conditions.  Therefore, because computers use finite-precision arithmetic,
individual runs are largely irreproducible from machine to machine.  In this context, one
can only make statements about the probability of a system behaving in a certain manner
(Quinlan \& Tremaine 1992). Thus, for each particular dimension of phase space explored, we
performed 300 runs, each
with randomly chosen initial orbital parameters that lay within specific ranges.

We expanded on the results of Ford, Havlickova, and Rasio (2001) by investigating the effects
of altering the planet/star mass ratio for each planet.  For each pair of mass ratios, 300 runs were
performed, each run with both planets having a randomly chosen initial eccentricity between 0 and
0.01 and initial inclination between $0^{\circ}$ and $5^{\circ}$, along with an argument of perihelion,
longitude of ascending node, and mean anomaly each randomly chosen between $0^{\circ}$ and $360^{\circ}$.

We performed 9 sets of 300 runs for each of the following four outer planet mass ratios $\mu_2$:
$1 \times 10^{-4},$ $5 \times 10^{-4},$ $1 \times 10^{-3},$ and $5 \times 10^{-3}.$  Each set of
runs corresponds to a different inner planet mass ratio. We varied the mass of the inner planet
within one order of magnitude of the mass of the outer planet. As mentioned, the initial
semimajor axis of the inner planet for every run was set at 1.0 AU.
The initial semimajor axis of the outer planet was chosen in a regime that exhibited unpredictable
behavior but allowed for quasi-periodic orbits.  Explanation of this regime follows.

\begin{figure*}
\centerline{\psfig{figure=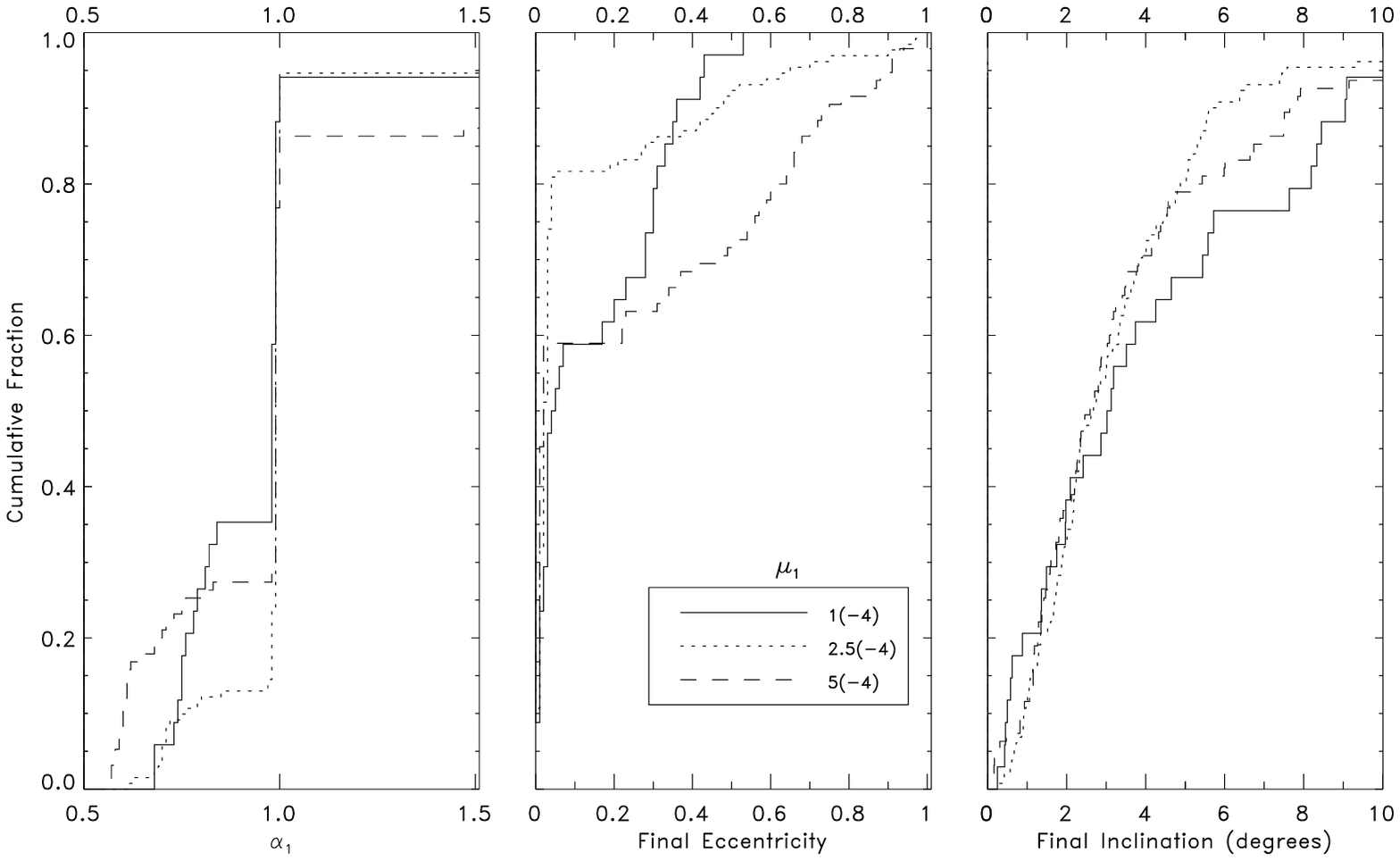,width=6.75truein,height=3.5truein}}
\vspace{-0.10truein}
\centerline{The Inner Planet}

\vbox{}
\vbox{}

\centerline{The Outer Planet}
\vspace{-0.35truein}
\centerline{\psfig{figure=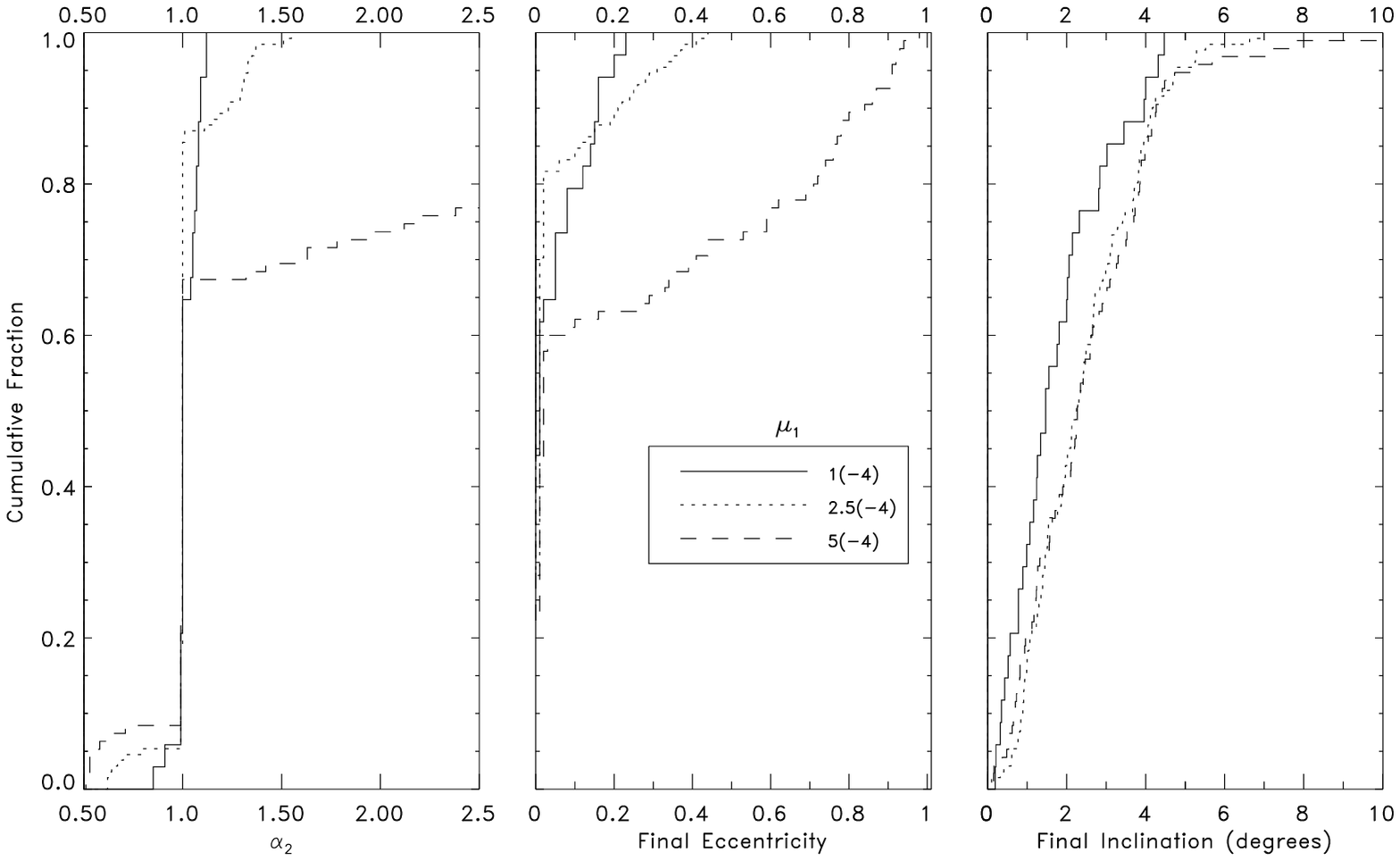,width=6.75truein,height=3.5truein}}

\caption{Cumulative probability distributions for stable systems with $\mu_2 = 5 \times 10^{-4}$.
For clarity, values of $\mu_1$ are abbreviated such that $1 \times 10^{-4} \equiv 1(-4)$.
The upper panel represents orbital parameter distributions of the inner planet, and the
lower panel the outer planet.}
\label{fig2}
\end{figure*}

\begin{figure*}

\centerline{\psfig{figure=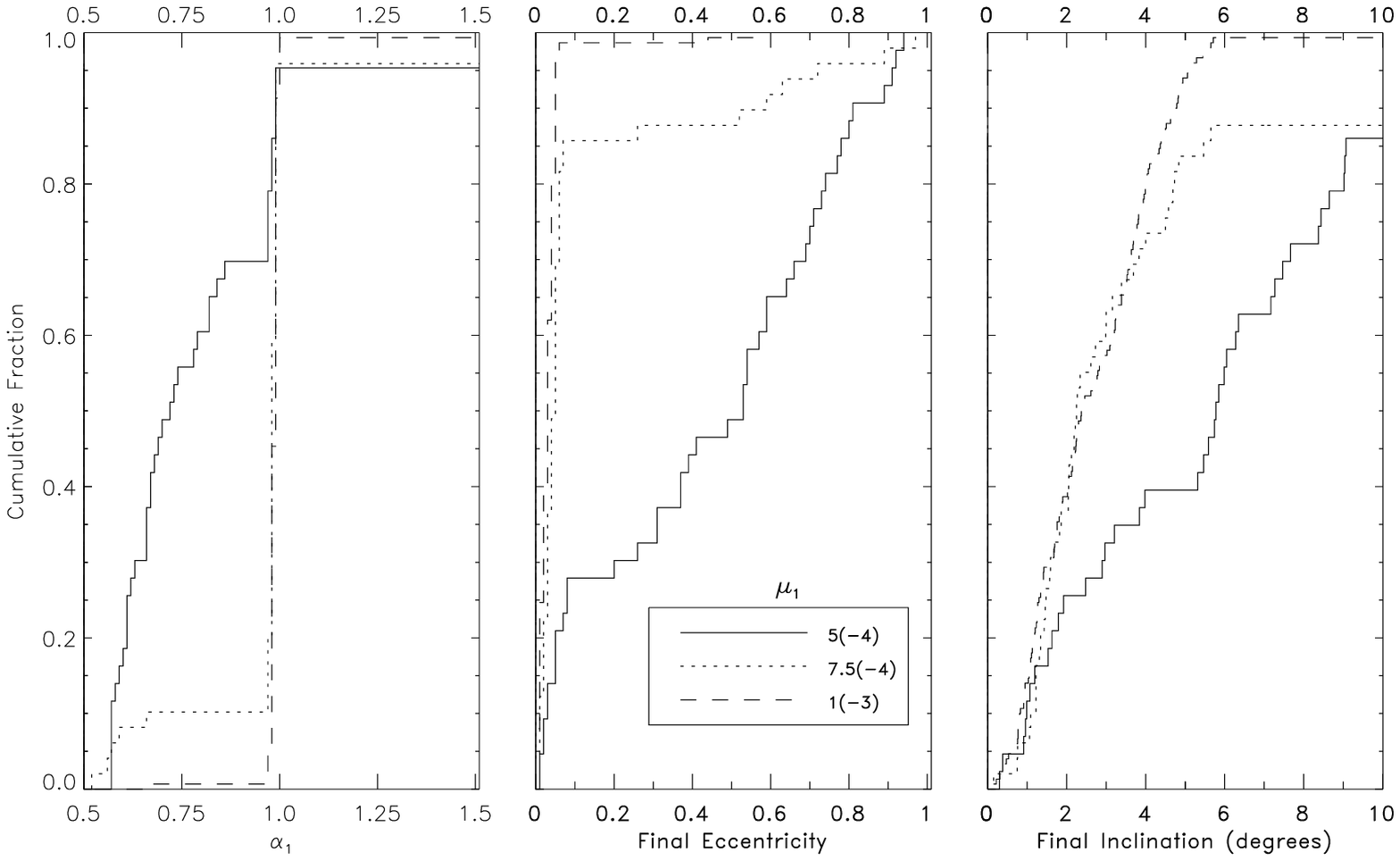,width=6.75truein,height=3.50truein}}
\vspace{-0.10truein}
\centerline{The Inner Planet}

\vbox{}
\vbox{}

\centerline{The Outer Planet}
\vspace{-0.35truein}
\centerline{\psfig{figure=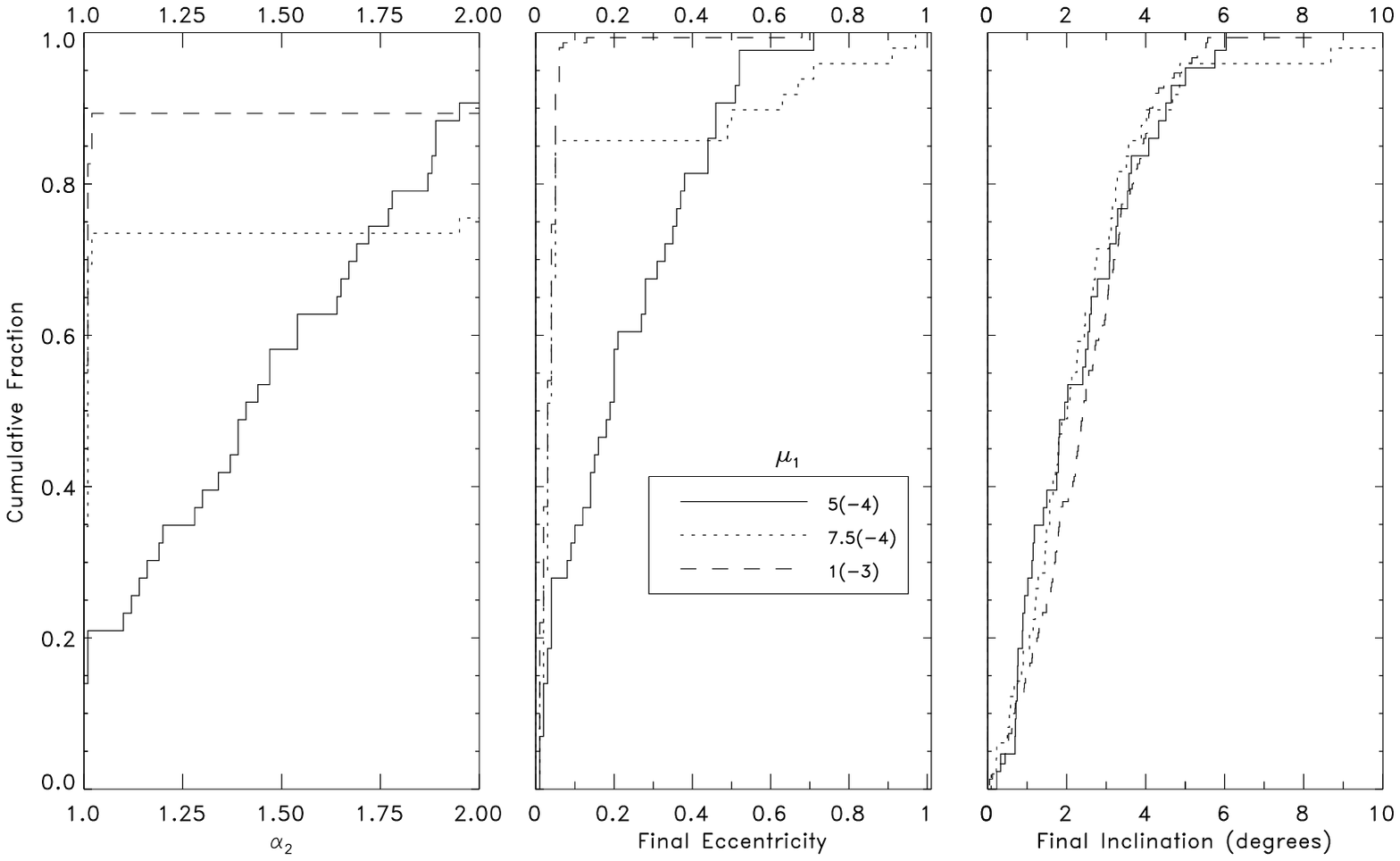,width=6.75truein,height=3.50truein}}

\caption{Cumulative probability distributions for stable systems with $\mu_2 = 1 \times 10^{-3}$.
For clarity, values of $\mu_1$ are abbreviated such that $5 \times 10^{-4} \equiv 5(-4)$.
The upper panel represents orbital parameter distributions of the inner planet, and the
lower panel the outer planet.}
\label{fig3}
\end{figure*}

\begin{figure*}
\centerline{\psfig{figure=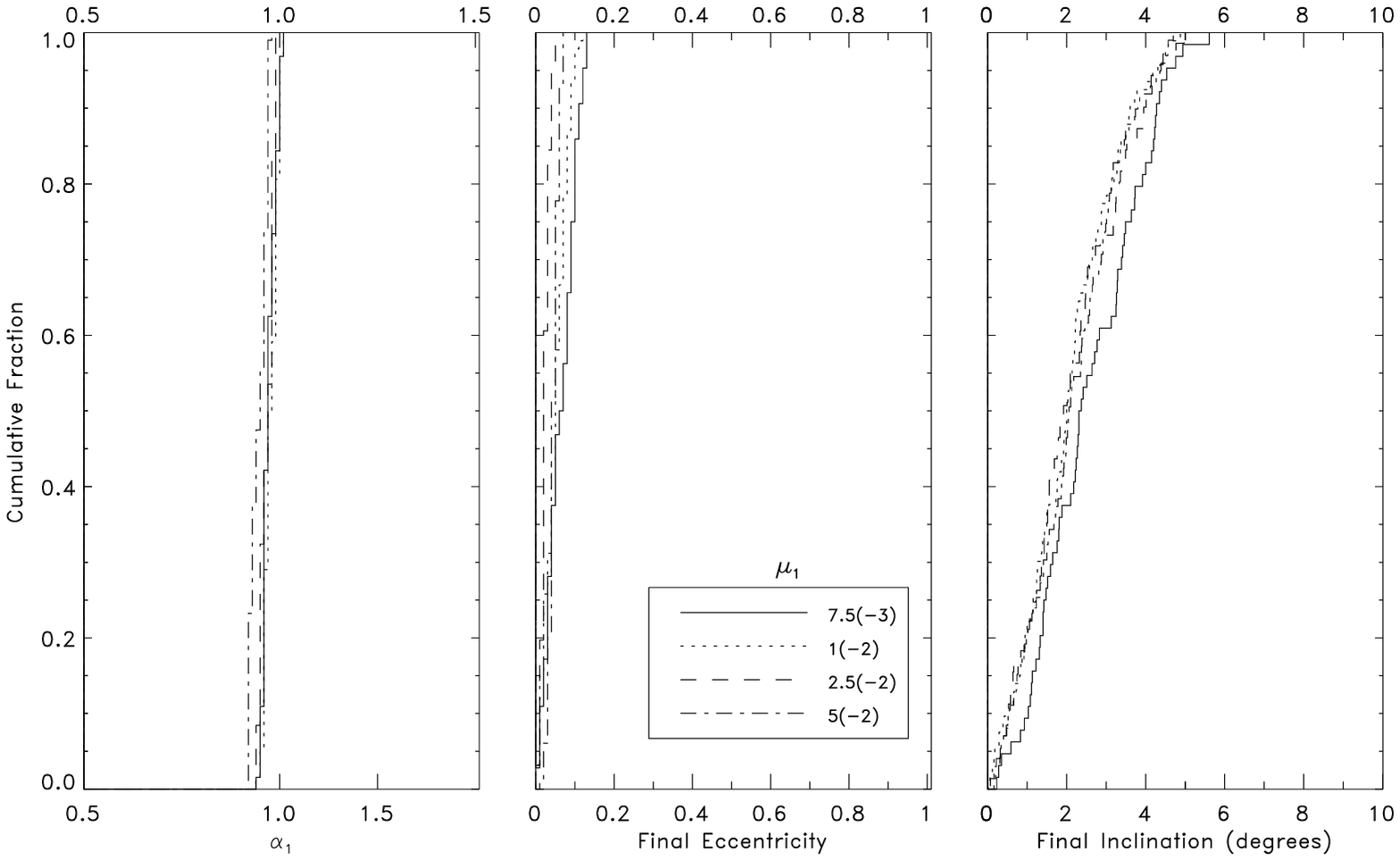,width=6.75truein,height=3.5truein}}
\vspace{-0.10truein}
\centerline{The Inner Planet}

\vbox{}
\vbox{}

\centerline{The Outer Planet}
\vspace{-0.35truein}
\centerline{\psfig{figure=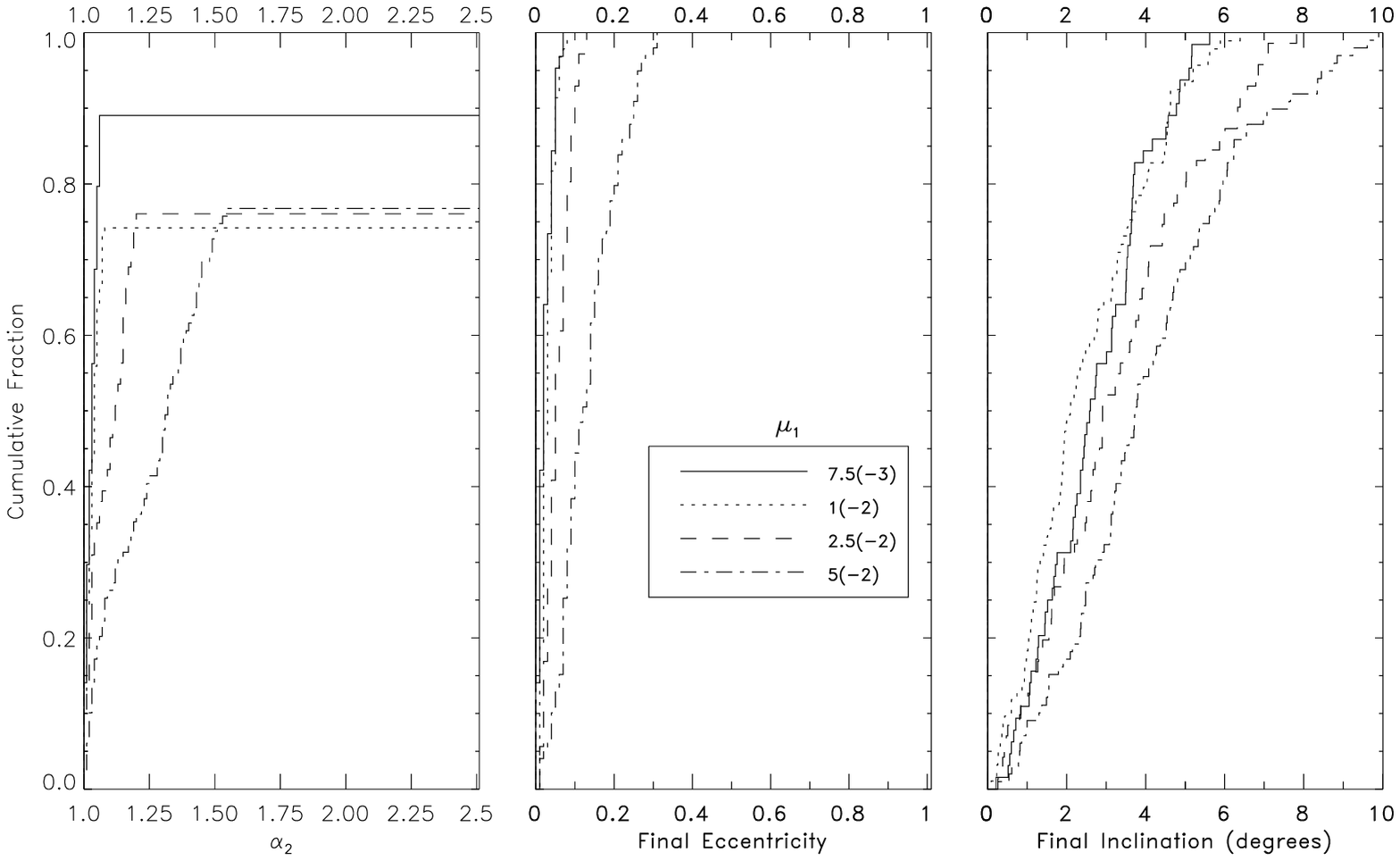,width=6.75truein,height=3.5truein}}

\caption{Cumulative probability distributions for stable systems with $\mu_2 = 5 \times 10^{-3}$.
For clarity, values of $\mu_1$ are abbreviated such that $7.5 \times 10^{-3} \equiv 7.5(-3)$.
The upper panel represents orbital parameter distributions of the inner planet, and the
lower panel the outer planet.}
\label{fig4}
\end{figure*}

\begin{figure*}
\centerline{\psfig{figure=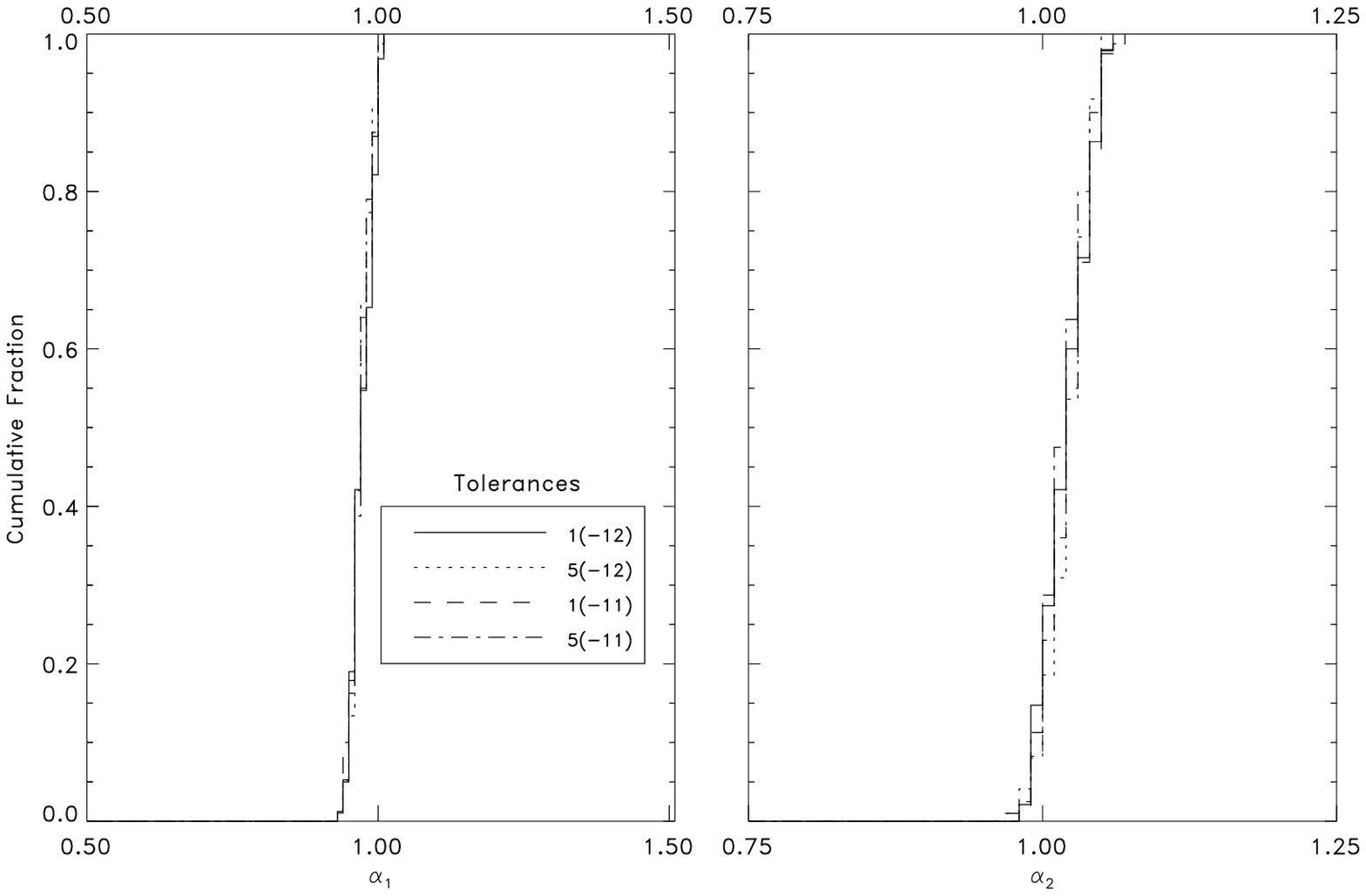,width=7.0truein,height=2.3truein}}

\vspace{-0.10truein}
\leftline{\ \ \ \ \ \ \ \ \ \ \ \ \ \ \ \ \ \ \ \ \ \ \ \ \ \ \
\ \ \ \ \ \ \ \ \ \ \ \ \ \
The Inner Planet \ \ \ \ \ \ \ \ \ \ \ \ \ \ \ \ \ \ \ \ \ \ \ \ \ \ \ 
\ \ \ \ \ \ \ \ \ \ \ \ \ \ \ \ \ \ \ \ \ \ \ \ \ \ \ \ \ \ 
The Outer Planet}

\vbox{}
\vbox{}

\leftline{\ \ \ \ \ \ \ \ \ \ \ \ \ \ \ \ \ \ \ \ \ \ \ \ \ \ \
\ \ \ \ \ \ \ \ \ \ \ \ \ \ 
The Inner Planet \ \ \ \ \ \ \ \ \ \ \ \ \ \ \ \ \ \ \ \ \ \ \ \ \ \ \ 
\ \ \ \ \ \ \ \ \ \ \ \ \ \ \ \ \ \ \ \ \ \ \ \ \ \ \ \ \ \   
The Outer Planet}
\vspace{-0.15truein}
\centerline{\psfig{figure=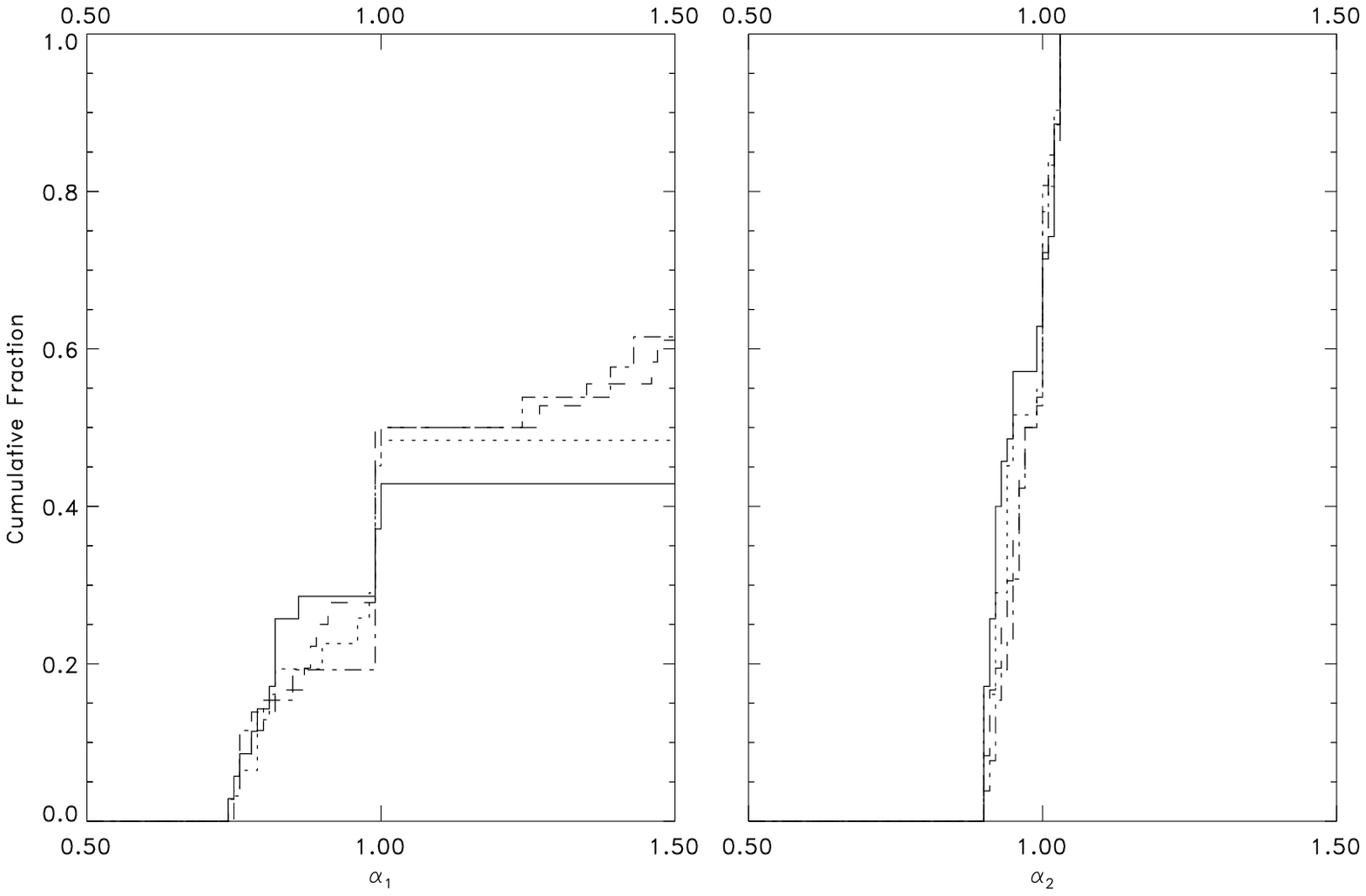,width=7.0truein,height=2.3truein}}

\caption{Cumulative probability distributions for a system with negligible migration ($\mu_1 =
\mu_2 = 5 \times 10^{-3}$, upper panel) and for a system with significant migration
($\mu_1 = 1 \times 10^{-5}, \ \mu_2 = 1 \times 10^{-4}$, lower panel) for different tolerances,
which are given with abbreviated scientific notation.  900 runs were performed for the upper
panel system, and 300 were performed for the lower panel system. Similarly to Figs.
\ref{fig1}-\ref{fig4}, only the stable systems are shown.}
\label{fig6}
\end{figure*}

As mentioned previously, the region bounded by $\Delta$ and $\Delta_{ch}$ describe systems
that $\it may$ exhibit Hill instability and $\it may$ exhibit quasi-periodic motion.
When both are exhibited, $\it some$ planets exhibit outward migration, which is the topic
of this section.  So as to include all regions from which migrating behavior might arise, we took
$\Delta_{ch} = 2 \cdot {[min(\mu_{1},\mu_{2})]}^{\frac{2}{7}}.$  Because $a_{1} = 1$ AU for
each run, values of $a_{2}$ were randomly chosen between $1 + \Delta_{ch}$ and
$1 + \Delta$ for the simulations.

Our initial conditions for the scattering experiments are similar to those adopted
by most previous authors (e.g. Ford, Havlickova \& Rasio 2001; Marzari \& Weidenschilling 2002;
Chambers, Wetherill \& Boss 1996), in that we start with fully-formed planets in relatively
close proximity to each other, and consider only their mutual gravitational interactions.
Such initial conditions are chosen primarily for simplicity and compatibility with previous
work, and can only partially be justified on physical grounds. In particular, since many of
the trial systems prove to be unstable over time scales comparable to the lifetime of
protoplanetary discs (Haisch, Lada \& Lada 2001), there is likely to be an earlier
epoch during which {\em both} planet-planet gravitational interactions and planet-disc
interactions are important.  This earlier stage of evolution could lead to an unstable
multiple planet system in at least two ways. First, two planets might form in well-separated
orbits, but then subsequently migrate into an unstable configuration as a consequence of
disc-driven migration (Goldreich \& Tremaine 1980; Lin \& Papaloizou 1986; Lin, Bodenheimer \&
Richardson 1996). For this to happen, the planets would have to avoid becoming trapped
into resonance during the migration process (Snellgrove, Papaloizou \& Nelson 2001;
Lee \& Peale 2002; Murray, Paskowitz \& Holman 2002).  Alternatively, the planets might form
close together, and be stabilized against immediate violent instability by the presence of
a surrounding gas disc (Lin \& Ida 1997; Nagasawa, Lin \& Ida 2003)\footnote{In related work,
Agnor \& Ward (2002) considered the damping of {\em terrestrial} planet eccentricity by a
remnant gas disc.}.  Further, unstable planetary systems may form after the disc has
dissipated.  Planet-planet interactions alone in crowded systems of Jovian-mass planets
most often leaves only two survivors in close, quasiperiodic but ultimately unstable orbits (Adams
\& Laughlin 2003). Although any of these pathways could plausibly lead to initial conditions
that are similar to those which we (and other authors) have assumed, a full treatment will
obviously need to model the more complex interactions that are possible during the phase
when the gas disc is being dispersed.

\subsection{Results}

Table 1 and Figs. \ref{fig1}-\ref{fig4} displays the data we took.  Table 1 summarizes the data
in terms of mass ratio and extent of migration, while the figures provide a more detailed look
at migration behavior.  Fig. \ref{fig5} provides a representative sample of the final semimajor
axes and eccentricities of both planets when significant outward migration occurs, and provides
a perspective on how data is plotted in Figs. \ref{fig1}-\ref{fig4}.  For example, the 43 stable
systems displayed in Fig. \ref{fig5} are plotted as the solid lines in Fig. \ref{fig3}.
Fig. \ref{fig5} illustrates that planets that migrate both inward and outward tend to
increase their eccentricity.  Although the extent of migration in both directions is similar,
those planets that migrate inward tend to increase their eccentricities at a higher rate.

Figs. \ref{fig1}-\ref{fig4} each represent a set of six cumulative
probability distributions that display the fraction of {\it stable} systems vs. semimajor axis ratios,
eccentricities, and inclinations of the outer and inner planet after exactly 2 Myr.  The upper panel in
each figure represents the initially outer planet, and the lower panel
represent the initially inner planet.  Each figure
keeps the mass of the outer planet fixed, but varies the mass of the inner planet.  Although
we obtained nine curves for each graph on each figure, we show only the curves that
exhibit significant radially outward migration, plus some that do not, and we do not show
any curves where less than 10\% of the 300 runs were stable.

Detecting migration for the outer planets is difficult because their
semimajor axes were randomly chosen for each run.  To aid in determining the net radial
movement of these planets, we define $\alpha_1 \equiv a_1({\rm final})/a_1({\rm initial})$  
and $\alpha_2 \equiv a_2({\rm final})/a_2({\rm initial})$.  Each panel in Figs.
\ref{fig1}-\ref{fig4} contain distributions of semimajor axis ratios, rather than
absolute semimajor axes, because those are scalable in this study.

Table 1 and Figs. \ref{fig1}-\ref{fig4} illustrate that outward migration of both the inner and
outer planets occurs more frequently with smaller values of $\mu_2$.  Fig. \ref{fig1} displays
extensive migration.  In the lowest mass case, almost 60\% of the stable inner planets migrate at
outward to at least 150\% of their original semimajor axis.  As $\mu_1$ is increased, 
migration occurs less frequently; for
$\mu_1 = 2.5 \times 10^{-4}$ no inner planets migrated out to 150\% of their original
semimajor axis.  
The correlation of mass ratio 
to final state behavior is less apparent but still present
in the final eccentricity and inclination curves.  In both graphs, the least-mass case 
prompts the highest
final eccentricity and inclination values for the orbit of the inner planet.

The final state of the outer planet, represented by the lower panel of Fig. \ref{fig1},
reflects the exchange of angular momentum in this system.  One sees that 
when an inner planet migrates outward, it leaves the
initially outward planet back so that $\alpha_2 < 1$.  The one instance where the outer
planet experiences significant migration is the highest mass case - the same case where
$\alpha_1 < 1.5$.  The eccentricity and inclination distributions for the outer planet in
this massive case differ drastically from the behavior seen for other inner planet masses.

In Fig. \ref{fig2}, $\mu_2 = 5 \times 10^{-4}$ and the result is significantly less 
migration than seen in
Fig. \ref{fig1}. Similarly to Fig. \ref{fig1}, the highest mass case exhibits the 
greatest migration of the outer planet.
In this case, $\simeq$20\% of the outer planets and $\simeq$15\% of the inner planets migrate
beyond 150\% of their original semimajor axis.  Halving the inner mass value reduces
the probability for $\alpha_1 < 1.5$ or $\alpha_2 < 1.5$ to $\simeq$5\%. In contrast
to Fig. \ref{fig1}, the least mass
case fails to show
significant outward migration of the inner planet, despite the inner planet exhibiting
the highest final inclinations.

In Fig. \ref{fig3}, where the outer mass is a full order of magnitude more massive than in
Fig. \ref{fig1}, we see
even less migration.  Inner bodies for the smallest mass case exhibit high eccentricities and
inclinations, but only $\simeq$5\% satisfy $\alpha_1 > 1.5$.  The case where
$\mu_1 = \mu_2 = 1 \times 10^{-3}$
is the same case studied by Ford, Havlickova, \& Rasio (2001).  Their Figs. 11-12 can be
compared to the semimajor axis ratio distributions, eccentricity distributions, and
inclination distributions in Fig. \ref{fig3}.

No migration of the inner planet occurs for any case in Fig. \ref{fig4}.  However, the outer
planet shows
significant migration.  Further, as shown by the sequence of four curves in the lower panel, 
the higher the mass of the outer planet, the greater the extent of the migration.  For the higher
mass cases, about 25\% of the initally outer planets 
satisfy $\alpha_2 > 2.5$.  Remarkably, these planets all retain final
eccentricities $< 0.4$, in stark
contrast to the large final eccentricities of the migrating planets of Figs.
\ref{fig1}-\ref{fig3}.

Because of the chaotic nature of the three body problem, one encounters difficulty
when predicting the appropriate timescale over which to integrate.  We adopted a similar
timescale to the one used by Ford, Havlickova, \& Rasio (2001), but recognize that systems
that appear to be stable at 2 Myr might
become unstable at some future time.  We extended the running time of one system to sample the
consequences.  After 2 Myr, 55 out of the 300 systems for the case $\mu_1 = 5 \times 10^{-3},
\ \mu_2 = 1 \times 10^{-3}$ remain stable. By running this system for 10 Myr, we found only
3 out of the 55 systems became unstable, and did so quickly.  Further, these three systems
became unstable before 3 Myr.  

For any set of 300 runs of a chaotic system, the slightest change in any input parameter of the 
Bulirsch-Stoer algorithm might drastically alter the results of any individual runs, but keep
the same global behavior.  The extent of the invariance of this global behavior is a function of
the code used and the number of systems sampled.  To explore this measure of invariance for
the runs presented here, we reran two systems with four different initial tolerances 
($5 \times 10^{-11}$, $1 \times 10^{-11}$, $5 \times 10^{-12}$, $1 \times 10^{-12}$).  One of
these systems ($\mu_1 = \mu_2 = 5 \times 10^{-3}$) exhibited negligible migration, and
the other ($\mu_1 = 1 \times 10^{-5}, \ \mu_2 = 1 \times 10^{-4}$) exhibited significant migration.
In order to achieve the largest feasible sample size for this error analysis, we tripled the
number of runs performed to 900 for the system exhibiting negligible migration.  As Fig. \ref{fig6}
illustrates, runs with different tolerances are practically indistinguishable for the case of no
migration, but vary up to 20\% for the case of significant migration.

\subsection{Summary}

The gravitational interaction of a pair of unequal mass planets that lie in close initial
configurations allow {\it either} planet to migrate outward to at least twice it's initial
semimajor axis.  Planets that migrate outward lie in quasi-stable, slightly inclined eccentric orbits
with an eccentricity that spans the entire permissible range and an inclination up to about
$10^{\circ}$.  Although either planet in any given system may drift outward, the smaller
the mass of the planets, the higher the tendency for the initially inner planet to migrate outward.
Further, less massive giant planets such as Uranus or Neptune are more likely to migrate
outward than Jupiter-mass planets.  This result, derived from our gravitational scattering
simulations alone, is consistent with Thommes, Duncan \& Levison's (1999, 2002) conclusion that
Uranus and Neptune's current location is a result of outward migration amongst Jupiter and Saturn.

\section{Conclusions}

As a first step in explaining the presence of planets at large orbital radii, we
have shown that outward migration of protoplanets is possible both by planet-disc
interactions and by planet-planet gravitational scattering without the presence of
a disc.  Strong mass loss in discs coupled with planetary cores that are formed
at about $\sim 10$ AU allow planets to migrate outward in discs to radii that are 
as much as a factor of several in excess of 
their initial semimajor axes. Planets that migrate in such a manner are likely
to be massive.  We predict that gas-driven outward migration should be most likely
to occur around more massive stars, whose strong UV flux can drive a powerful
photoevaporative outflow. Subsequently, when in the appropriate chaotic regime,
planets within a multiple planet system may migrate outward due to gravitational
scattering alone.  Planets that migrate in this manner may be massive or not, however
low mass objects tend to exhibit the most extensive outward migration. Orbital migration
is typically accompanied by an increase in eccentricity that spans the allowable
range for elliptic orbits.

\section*{Acknowledgments}

We thank an anonymous referree for helpful comments, Kevin Rauch and Doug Hamilton
for use of their integrator and assistance in it's operation, as well as the
Colorado Rings Group for useful discussions.

This paper is based upon work supported by NASA under Grant NAG5-13207 issued 
through the Office of Space Science.

\end{document}